\documentclass[10pt]{iopart}
\usepackage{iopams}
\usepackage{amssymb}
\usepackage[dvips]{graphicx}
\date{\today}

\usepackage{epsfig}
\usepackage{bbm}

\newcommand{\1}{\mathbbm{1}}
\renewcommand{\vec}[1]{\mbox{\boldmath $#1 $}}
\begin{document}
\title{Learning by a neural net in a noisy environment -- The pseudo-inverse 
solution revisited}
\author{W A van Leeuwen and B Wemmenhove\footnote[1]{Correspondence should be 
addressed to wemmenho@science.uva.nl}}
\address{Institute for Theoretical Physics, University of Amsterdam,
Valckenierstraat 65, 1018 XE Amsterdam, 
The Netherlands}

\begin{abstract}
A recurrent neural net is described that learns a set of patterns 
$\{\vec{\xi}^{\mu}\}$ in the
presence of noise. The learning rule is of a Hebbian type, and, if noise 
would be absent during the learning process, the resulting final values of the
weights $w_{ij}$ would correspond to the pseudo-inverse solution of the
fixed point equation in question. 
For a non-vanishing noise parameter, an explicit expression for the expectation
value of the weights is obtained. This result turns out to be 
unequal to the pseudo-inverse solution.
Furthermore, the stability properties of the system are discussed.

\end{abstract}
\submitto{\JPA}

\maketitle

\section{Introduction and summary}

In principle, it is our purpose to study learning in a neural net as it 
occurs in nature. The theory of recurrent neural nets \cite{MRS95} provides 
us with a model of content-addressable memory as it might be realized, to 
some extent, in the brain. 
Learning, in such a model, corresponds to adjusting the synaptic matrix
$w_{ij}$ in such a way that $p$ memorized patterns $\vec{\xi}^{\mu}$,
$(\mu=1, \ldots, p)$, become fixed points of the neuron state dynamics.
This can be achieved in a recurrent neural net by sequentially 
clamping its neurons to a well-defined and unique set of patterns, and 
adjusting
the weights of the connections according to some Hebbian learning rule. 
However, in reality, a neural net cannot be clamped to a fixed set of ideal 
patterns. 
A more realistic assumption would be 
that the clamping of the net to a pattern always is more or less
distorted. 
Consider, for instance, the visual system as a system in which the clamping
is imposed by input from the retina. Since neurons are noisy objects, which
once in a while fire spontaneously, an internal representation of a stimulus
in the brain will hardly ever be identical to the
representation corresponding to a previous stimulus.

We therefore introduce noise to the set of patterns, thus making the set 
less well-defined and less unique.
A network state array of a net of $N$ neurons is denoted by 
\begin{equation}
\vec{x} := (x_1, \ldots, x_N) \label{vecx}
\end{equation}
where $x_i=1$ if neuron $i$ is active and $x_i=0$ if it is non-active.
At every learning step $n$, $\vec{x}$ will be similar to one
of the $p$ given patterns $\vec{\xi}^{1}, \ldots, \vec{\xi}^p$, but it has 
nonzero probability, for each neuron $i$, $(i=1, \ldots, N)$, of deviating 
from it.
At each learning step $n$, 
synaptic connections $w_{ij}$ will adapt themselves, according to a Hebbian 
learning rule which
is a function of the weights $w_{ik}$ and of the 
(binary) neuron states $x_k$, $(k=1, \ldots, N)$. 
For the class of learning rules that we will use in our model, the case of 
noiseless learning has been studied in detail 
(\cite{PGD85}, \cite{DieOp}, \cite{Hee99}). 
At every learning step $n$ a pattern $\vec{\xi}^{\mu}$, $(\mu=1, \ldots p)$ is
chosen. 
If, for each $n$, we put 
$\vec{x} = \vec{\xi}^{\mu}$, the resulting weights $w_{ij}$ for 
$n\rightarrow \infty$ are known to coincide with the pseudo-inverse solution. 
The pseudo-inverse solution is a particular solution of the under-determined
set of $pN$ equations
\begin{equation}
\displaystyle{\sum_{j=1}^N} w_{ij} \xi_j^{\mu} -\theta_i
= \kappa(2\xi_i^{\mu}-1)
\label{PIequations}
\end{equation}
for the $N(N-1)$ unknowns $w_{ij}$ $(i,j=1,\ldots , N; i \neq j)$, $p<N$.
Here, $\kappa$ is a positive number, and $\theta_i$ a constant.
It is easily verified that these equations guarantee that the so-called 
stability coefficients 
\begin{equation}
\gamma_i(\vec{\xi}^{\mu}) = (2\xi_i^{\mu}-1)(\displaystyle{\sum_{j=1}^N} 
w_{ij} \xi_j^{\mu} -\theta_i)
\label{stabcoef}
\end{equation}
are positive for all patterns
$\vec{\xi}^{\mu}$. In our description, however, 
the network state during learning is determined by a probability distribution 
$p^{\mu}(\vec{x})$, centered around
the patterns $\vec{\xi}^{\mu}$.
By means of the Master equation derived in section \ref{mastersec}, we 
will arrive at the following 
equation for the expectation value of the weights in the limit of 
$n\rightarrow \infty$:
\begin{equation}
\frac{1}{p}\displaystyle{\sum_{\mu=1}^p \sum_{\vec{x} \in \Omega}}
p^{\mu}(\vec{x})
[\kappa(2x_i -1) - \bigl(\displaystyle{\sum_{k=1}^N}
\langle w_{ik} \rangle_{\infty} x_k -\theta_i\bigr)]x_j = 0
\label{Ourequation}
\end{equation}
where $\Omega$ is the collection of all possible $2^N$ arrays $\vec{x}$. 
In contrast to the equations (\ref{PIequations}), this is a completely 
determined set of equations, of which the solution 
is essentially different from the pseudo-inverse solution of 
(\ref{PIequations}). It will turn out to exist only 
if the variance of the probability distribution $p^{\mu}(\vec{x})$ is non-zero,
\emph{i.e.}, in the presence of noise. This is the subject of section 
\ref{solutionsec}.

In section \ref{intermediatesec} we will study numerically the weights 
$\langle w_{ij} \rangle_n$ as a function of $n$. 

In section \ref{stabilitysection} we will show that, on the
average, this solution does yield stability coefficients close to
$\kappa$ if the `noise parameter', $b$, which will be introduced
in the probability distribution $p^{\mu}(\vec{x})$, is small enough, 
which shows that the solution found is stable indeed.

Finally, in section \ref{basinssection}, we study the size of the basins of 
attraction around our new solution. It is found 
that the sizes are larger than those around the pseudo-inverse solution, a 
result that is in perfect agreement with earlier observations 
\cite{GSW89, WS90a, WS90b, WS93} that learning with noise enlarges the basins
of attraction. In these earlier studies, however, no analytical expression 
for the average values of the weights has been given.

\section{Derivation of the Master Equation for a linear learning rule}
\label{mastersec}
We consider a recurrent net of $N$ binary neurons. 
The strengths of the synaptic 
connection between the post-synaptic neuron $j$ and the pre-synaptic 
neuron $i$ will be denoted by $w_{ij}$. The neurons $i$ $(i=1,\ldots, N)$ can 
take the values 
$x_i=0$ or $x_i=1$, corresponding to the non-active and active state, 
respectively. It is useful to associate with each neuron $i$ a set $V_i$, 
defined as the collection of neuron indices $j$ with which neuron $i$ has an 
adaptable, \emph{i.e.}, a non-zero, non-constant afferent synaptic connection.
In other words, for all $j \in V_i$, there is an axon going from neuron 
$j$ to a dendrite of neuron $i$, and the corresponding weight $w_{ij}$ is 
adaptable in a learning process.
The collection of neurons $j$ with which $i$ has no connection, or a 
non-changing synaptic connection, will be denoted as the complementary set,
$V_i^C$.
We suppose that the synaptic strengths $w_{ij}$ between the neurons are 
changed in steps according to a rule of 
the general form 
\begin{eqnarray}
w_{ij}' =  w_{ij}  +  \Delta w_{ij}   \qquad& j \in V_i \\
w_{ij}'  =  w_{ij}    & j \in V_i^C
\label{leerregel}
\end{eqnarray}
where $\Delta w_{ij}$ is a function of the states $x_k$ of all $N$ neurons of
the net and all afferent synaptic weights $w_{ik}$ ($i$ fixed, 
$k=1,2,\ldots,N)$. In general, the functions $\Delta w_{ij}$ will be 
linear in all $x_k$,
since $x_k^2 = x_k$  (recall that $x_k$ equals $0$ or $1$), but non-linear in 
the weights $w_{ik}$. In
 this article we suppose, however, that the $\Delta w_{ij}$ do depend linearly 
on the weights $w_{ik}$. Hence, in this article, 
\begin{equation}
\Delta w_{ij} = \Delta w_{ij} (\vec{x},\vec{w}_i)
\end{equation}
is a linear function in all $x_k$ and all $w_{ik}$ ($k=1,2,\ldots N$). We
abbreviated
\begin{equation}
\vec{w}_i := (w_{i1}, \ldots, w_{iN}) \label{vecwi}
\end{equation}
It is unrealistic to describe a biological neural net as a 
deterministic system, since there are many unknown parameters 
that influence its development in time. We therefore choose a 
probabilistic description. We suppose that the neuron states $x_i$ are 
mutually independent stochastic variables, \emph{i.e.}, 
the probability that neuron
$i$ has value $x_i$ is given by a probability distribution $p_i(x_i)$
which is independent of $j$ $(j\neq i)$. Since
the changes $\Delta w_{ij}$ of the weights $w_{ij}$ are functions of the 
stochastic variables $x_i$ $(i=1, 2, \ldots N)$, and a function of a 
stochastic variable is a 
stochastic variable, the changes $\Delta w_{ij}$, and, hence, the 
$w_{ij}$ themselves, are stochastic variables. 

Now let $T_{ij}(w_{ij}'|w_{ij},\{ w_{ik} \}_{k\neq j})$ be the probability 
that, due to a learning 
step, a transition takes place from the value $w_{ij}$ to the value 
$w_{ij}' = w_{ij} + \Delta w_{ij}$, for a given set $\{ w_{ik} \}_{k\neq j}$.
Then we have
\begin{equation}
T_{ij}(w_{ij}'|w_{ij},\{w_{ik}\}_{k\neq j}) = 
\displaystyle{\sum_{\vec{x} \in \Omega}} p(\vec{x})
\delta(w_{ij}'-w_{ij}- \Delta w_{ij}(\vec{x}, \vec{w}_i))
\label{transition}
\end{equation}
where $\Omega$ is the collection of all $2^N$ possible states of the neural 
net $(x_i=0,1;i=1, \ldots, N)$ and
$p(\vec{x})$
is the probability of occurrence of the network state $\vec{x}$, which
we suppose to be independent of the variables $w_{ij}$. [The relation between
$p(\vec{x})$ and $p_i(x_i)$ is left unspecified at this stage of the 
reasoning; compare, however, (\ref{pxvoorp}) and (\ref{pmufactor}) below.]
We have
\begin{equation}
\displaystyle{\sum_{\vec{x} \in \Omega}} p(\vec{x}) = 1
\label{somp(x)}
\end{equation}
The delta-function in (\ref{transition}) guarantees that only transitions 
take place which obey
the learning rule (\ref{leerregel}). Using (\ref{somp(x)}) we find from
(\ref{transition}), that 
\begin{equation}
\int T_{ij}(w_{ij}'|w_{ij},\{w_{ik}\}_{k\neq j})dw_{ij}' = 1 
\label{normTij}
\end{equation}
Let $P_{ij}(w_{ij},n)$ be the probability of occurrence of the 
variable $w_{ij}$ at a time step $n$ $(n=0,1,2,\ldots)$. Then $P_{ij}$ and 
$T_{ij}$ are related 
according to 
\begin{equation}
P_{ij}(w_{ij}, n+1) = \int \ldots \int T_{ij}(w_{ij}|\{w_{ik}'\})
\displaystyle{\prod_{k=1}^N}[P_{ik}(w_{ik}',n) dw_{ik}']
\label{Pij}
\end{equation}
Demanding that the probability $P_{ij}$ is normalized initially,
\begin{equation}
\int P_{ij}(w_{ij}, 0) dw_{ij} = 1
\label{normPij0}
\end{equation}
we find from (\ref{Pij}) and (\ref{normTij}), by induction, that 
\begin{equation}
\int P_{ij}(w_{ij},n) dw_{ij} = 1
\label{normPijn}
\end{equation}
for all $n$.
From (\ref{normTij}) and (\ref{Pij}) it follows that 
\begin{eqnarray}
\fl P_{ij}(w_{ij},n+1) - P_{ij}(w_{ij}, n) = 
\int \ldots \int [T_{ij}(w_{ij}|\{w_{ik}'\}) P_{ij}(w_{ij}',n) \nonumber \\
 - T_{ij}(w_{ij}'|w_{ij},\{w_{ik}'\}_{k\neq j}) P_{ij}
(w_{ij},n)]\displaystyle{\prod_{k\neq j}} [P_{ik}(w_{ik}',n)dw_{ik}']dw_{ij}' 
\nonumber \\ \qquad \qquad (i=1,\ldots, N; j \in V_i)
\label{master}
\end{eqnarray}
which is the so-called Discrete Master Equation for the weights $w_{ij}$.
It masters the evolution of the weights $w_{ij}$ as a function of $n$, and
determines the values of the weights in the long run. 

In order to obtain an expression for the expectation value of the weights
after infinitely many learning steps, we first consider the
expectation value at time step $n$:
\begin{equation}
\langle w_{ij} \rangle_n := \int P_{ij}(w_{ij},n) w_{ij}dw_{ij} \qquad 
(j\in V_i)
\label{expvalue1}
\end{equation}
The latter expression yields, using the Master Equation (\ref{master}),
\begin{eqnarray}
\fl \langle w_{ij} \rangle_{n+1} - \langle w_{ij} \rangle_n =
\int \ldots \int 
w_{ij} [T_{ij}(w_{ij}|\{w_{ik}'\}) P_{ij}(w_{ij}',n) \nonumber \\
 - T_{ij}(w_{ij}'|w_{ij},\{w_{ik}'\}_{k\neq j}) P_{ij}
(w_{ij},n)]\displaystyle{\prod_{k\neq j}} [P_{ik}(w_{ik}',n)dw_{ik}']dw_{ij}'
dw_{ij}
\label{masterw1}
\end{eqnarray}
or, interchanging the primed and unprimed variables $w_{ij}$ and
$w_{ij}'$ in the first term on the right hand side,
\begin{eqnarray}
\fl \langle w_{ij} \rangle_{n+1} - \langle w_{ij} \rangle_n =
\int \ldots \int (w_{ij}'-w_{ij}) T_{ij}(w_{ij}'|w_{ij},\{w_{ik}'\}_{k\neq j}) 
P_{ij}(w_{ij},n) \nonumber \\
\times \displaystyle{\prod_{k\neq j}} [P_{ik}(w_{ik}',n)dw_{ik}']dw_{ij}'
dw_{ij}
\label{masterw2}
\end{eqnarray}
or, with (\ref{transition}) and integrating over $w_{ij}'$,
\begin{equation}
\fl \langle w_{ij} \rangle_{n+1} - \langle w_{ij} \rangle_n =
\displaystyle{\sum_{\vec{x} \in \Omega}} p(\vec{x}) \int \ldots \int 
\Delta w_{ij}(\vec{x}, \vec{w}_i)
\displaystyle{\prod_{k=1}^N}  P_{ik}(w_{ik},n) dw_{ik} \qquad (j \in V_i)
\label{masterw3}
\end{equation}
or, with (\ref{normPijn}) and (\ref{expvalue1}),
\begin{equation}
\langle w_{ij} \rangle_{n+1} - \langle w_{ij} \rangle_n =
\displaystyle{\sum_{\vec{x} \in \Omega}} p(\vec{x})\Delta w_{ij}(\vec{x}, 
\langle \vec{w}_i \rangle_n), \qquad (j \in V_i)
\label{masterw4}
\end{equation}
where we used that $\Delta w_{ij}$ is linear in the $w_{ik}\ (k=1,\ldots, N)$
to replace $\vec{w}_i$ by the expectation value $\langle \vec{w}_i \rangle_n$
in the expression for $\Delta w_{ij}$.
If we assume that the expectation values of the synaptic connections 
$\langle w_{ij} \rangle_n$ converge to finite values, 
$\langle w_{ij} \rangle_{\infty}$, for $n$ tending to infinity, we can 
solve this equation for $n\rightarrow \infty$. This is the 
subject of the next section.

\section{Final values for the weights}
\label{solutionsec}
If we suppose that the left-hand side of (\ref{masterw4}) vanishes in the limit
of $n$ tending to infinity, we have 
\begin{equation}
\displaystyle{\sum_{\vec{x} \in \Omega}} p(\vec{x}) \Delta w_{ij}(\vec{x}, \langle
\vec{w}_i \rangle_{\infty}) = 0
\label{masterw5}
\end{equation}
At this point, we need an expression for the increment $\Delta w_{ij}(n)$, in
the $n$-th learning step. 
We take the biologically motivated learning rule \cite{Hee99}
\begin{equation}
\Delta w_{ij}(n) = \eta_i [\kappa - \gamma_i(\vec{x},n)](2x_i-1)x_j
\qquad (i=1, \ldots, N; j \in V_i)
\label{leerregelHee}
\end{equation}
where $\eta_i$ is the learning rate, $\kappa$ the margin 
parameter and $\gamma_i(\vec{x}, n)$ the stability coefficient given by
\begin{equation}
\gamma_i(\vec{x},n) = (2x_i -1)[h_i(\vec{x},n) -\theta_i]
\label{stabiliteit}
\end{equation}
[cf. eq. (\ref{stabcoef})].
Here, $h_i(\vec{x},n)$ is the membrane potential of neuron $i$ at step $n$ of 
the learning process
\begin{equation}
h_i(\vec{x}, n) = \displaystyle{\sum_{k=1}^N} w_{ik}(n) x_k 
\label{potential}
\end{equation}
and $\theta_i$ the threshold potential of neuron $i$. It should be noted that
in (\ref{potential}) $x_k$ is the state of neuron $k$ at step $n$ of the 
learning procedure.
Substituting (\ref{leerregelHee}) with (\ref{stabiliteit}) and 
(\ref{potential})
into (\ref{masterw5}) we find, using the fact that $(2x_i-1)^2=1$,
\begin{equation}
\displaystyle{\sum_{\vec{x} \in \Omega}}p(\vec{x})[\kappa(2x_i -1) - 
\bigl(\displaystyle{\sum_{k=1}^N}
\langle w_{ik} \rangle_{\infty} x_k -\theta_i\bigr)]x_j = 0
\label{nlarge1}
\end{equation}
where we divided by the learning rate $\eta_i$. 

Up to now, the precise form of the probability distribution $p(\vec{x})$ has
been left unspecified. At this point, let us specify our probability 
distribution $p(\vec{x})$ to be such that the chosen patterns $\vec{x}$ are  
centered around representative patterns $\vec{\xi}^{\mu}$.
To that end, we choose our probability distribution $p(\vec{x})$ such 
that it is a sum of $p$ equally probable, individually independent 
probability distributions, \emph{i.e.},
\begin{equation}
p(\vec{x}) = \frac{1}{p}\displaystyle{\sum_{\mu = 1}^p} p^{\mu}(\vec{x})
\label{pxvoorp}
\end{equation}
where $p^{\mu}(\vec{x})$ is factorizable,
\begin{equation}
p^{\mu}(\vec{x}) = \displaystyle{\prod_{i=1}^N}p^{\mu}_i(x_i)
\label{pmufactor}
\end{equation}
\emph{i.e.}, the neurons behave independently from one another. The quantity
$p_i^{\mu}(x_i)$ is the probability that, once the pattern index $\mu$
is chosen, neuron $i$ is in the state $x_i$. One therefore has 
\begin{equation}
p_i^{\mu}(0) + p_i^{\mu}(1) = 1
\label{imunorm}
\end{equation}
In a learning process, at every step $n$, the index $\mu$ is drawn from a
collection of $p$ equally probable pattern indices, thus fixing the 
probability distribution $p^{\mu}(\vec{x})$ 
according to which the pattern $\vec{x}$ is chosen for that learning step $n$.

Let us denote averages with respect to the probability $p^{\mu}(\vec{x})$
by $\overline{\phantom{x_i}}^{\mu}$
\begin{eqnarray}
\displaystyle{\sum_{x_i=0,1}} p^{\mu}_i(x_i) x_i = \overline{x_i}^{\mu}, \qquad
&\displaystyle{\sum_{x_i=0,1}} p^{\mu}_i(x_i) x_i^2 = \overline{(x_i^2)}^{\mu} 
\label{average1}
\end{eqnarray}
implying, in view of (\ref{pmufactor}) and (\ref{imunorm}),
\begin{eqnarray}
\displaystyle{\sum_{\vec{x} \in \Omega}} p^{\mu}(\vec{x}) x_i = 
\overline{x_i}^{\mu}, \qquad &
\displaystyle{\sum_{\vec{x} \in \Omega}} p^{\mu}(\vec{x}) x_i^2 = 
\overline{(x_i^2)}^{\mu} 
\label{average2}
\end{eqnarray}
Thus a bar with an index $\mu$ indicates an average with respect to the 
probability distribution $p^{\mu}(\vec{x})$.
With the choice (\ref{pxvoorp}), the result (\ref{nlarge1}) can be rewritten 
in terms of these averages, where we must take be aware that a term 
$\overline{(x_j^2)}^{\mu}$ appears in the
sum over $k$:
\begin{eqnarray}
\fl \langle w_{ij} \rangle_{\infty} \displaystyle{\sum_{\mu=1}^p} 
\left[\overline{(x_j^2)}^{\mu}
- \left(\overline{x_j}^{\mu}\right)^2 \right] =    
\displaystyle{\sum_{\mu =1}^p} 
\left[\kappa(2\overline{x_i}^{\mu} -1) - (\displaystyle{\sum_{k=1}^N} 
\langle w_{ik} \rangle_{\infty} \overline{x_k}^{\mu} - \theta_i)\right]
\overline{x_j}^{\mu} \ \ \ j \in V_i
\label{nlarge2}
\end{eqnarray}
The latter result can be rewritten as
\begin{equation}
p\sigma^2_j \langle w_{ij} \rangle_{\infty} 
= -\displaystyle{\sum_{k \in V_i}} (A_i)_{jk} \langle w_{ik} \rangle_{\infty} 
+ B_{ij}, \ \ 
j \in V_i
\label{nlarge3}
\end{equation}
where we abbreviated
\begin{eqnarray}
\lo{\sigma^2_j}  = \frac{1}{p}\displaystyle{\sum_{\mu=1}^p}
\overline{(x_j-\overline{x_j}^{\mu})^2}^{\mu} \nonumber \\
 =  \frac{1}{p} \displaystyle{\sum_{\mu=1}^p} \left[\overline{(x_j^2)}^{\mu} - 
\left(\overline{x_j}^{\mu}\right)^2 \right]
\label{variance}
\end{eqnarray}
and where we split up the sum over all $k$ in a sum over $V_i$ and a sum over 
its complement $V_i^C$:
\begin{eqnarray}
\fl (A_i)_{jk} :=  \displaystyle{\sum_{\mu=1}^p} \overline{x_j}^{\mu} 
\overline{x_k}^{\mu},  & i=1,\ldots, N; \ \ j,k \in V_i
\label{Aij}
\\
\fl B_{ij} :=  \displaystyle{\sum_{\mu=1}^p} 
[ \kappa (2\overline{x_i}^{\mu}-1)-
(\displaystyle{\sum_{k \in V_i^C}} \langle w_{ik} 
\rangle_0\overline{x_k}^{\mu} - \theta_i)]\,
\overline{x_j}^{\mu},  \qquad &i=1,\ldots, N; \ \ j \in V_i
\label{Bij} 
\end{eqnarray}
Note that the matrix $A_i$ is a symmetric matrix, the dimension of which 
equals the number of indices in $V_i$, \emph{i.e.}, the number of adaptable 
afferent 
synaptic connections of neuron $i$.
In the matrix $B$, we could write $\langle w_{ik} \rangle_0$ rather than 
$\langle w_{ik} \rangle_{\infty}$, since 
$\langle w_{ik} \rangle_0 = \langle w_{ik} \rangle_{\infty}$ for $k \in V_i^C$.
It is easy to solve the equation (\ref{nlarge3}). First, rewrite it as
\begin{equation}
\displaystyle{\sum_{k \in V_i}} \left[(D_i)_{jk} + (A_i)_{jk}\right] \langle w_{ik} 
\rangle_{\infty} = B_{ij}
\label{nlarge4}
\end{equation}
where $D_i$ is the diagonal matrix 
\begin{equation}
(D_i)_{jk} := p\sigma^2_j \delta_{jk}, \qquad j,k \in V_i
\label{Dij}
\end{equation}
The matrix $D_i + A_i$ is non-singular, and can be inverted. Inserting the
explicit form of $B_{ij}$ (\ref{Bij}), we then find
\begin{equation}
\fl \langle w_{ij} \rangle_{\infty} = \displaystyle{\sum_{k \in V_i}} (D_i + A_i)^{-1}_{jk}\displaystyle{\sum_{\mu=1}^p} [ \kappa (2\overline{x_i}^{\mu}-1)-
(\displaystyle{\sum_{l \in V_i^C}} \langle w_{il} 
\rangle_0\overline{x_l}^{\mu} - \theta_i)]\,
\overline{x_k}^{\mu} \ \ \ j \in V_i
\label{solutionwij}
\end{equation}
where we used that $D_i$ and $A_i$ are symmetric matrices. 
In the usual treatments of noiseless recurrent neural networks 
($\sigma_j=0$ for all $j$), one finds for the $w_{ij} (\infty)$
the so-called pseudo-inverse solution \cite{DieOp}, \cite{Hee99}, which 
reads, in our notation,
\begin{equation}
\fl w_{ij}^{PI} = w_{ij}(0) + \displaystyle{\sum_{\nu, \mu =1}^p}
(C_i^{-1})^{\mu \nu}[\kappa(2\xi_i^{\mu} -1) - (\displaystyle{\sum_{k=1}^N}
w_{ik}(0)\xi_k^{\mu} - \theta_i)]\xi_j^{\nu} \ \ \ j \in V_i
\label{PI}
\end{equation}
where $C_i^{-1}$ is the inverse of the correlation matrix
$C_i^{\mu \nu} = \sum_{k \in V_i} \xi_k^{\mu} \xi_k^{\nu}$.
Apparently, our result (\ref{solutionwij}) is not a simple generalization of
 the 
standard result for noiseless recurrent neural networks.
Note that the usual pseudo-inverse solution (\ref{PI}) depends on the 
initial values $w_{ij}(0)$ of all the weights, whereas our solution 
(\ref{solutionwij}) depends only on $w_{ij}(0)$ for $j \in V_i^C$ and not on
the initial value $w_{ij}(0)$ of the changing weights ($j \in V_i$). 
Apparently, a little bit of noise 
completely wipes out the effect of the initial state of changing
connections, since the result 
(\ref{solutionwij}) is true for any value of the noise unequal zero. 

In the limit that all $\sigma_j$ (\ref{variance}) vanish, the set 
of equations (\ref{nlarge4}) becomes under-determined for
$p<N$, since the matrix
$A_i$ is then singular. Hence, the solution 
(\ref{solutionwij}) does not exist for a noiseless net. Explicitly, this can 
be seen as follows.
Let us suppose that the $p$ average patterns $\{ \overline{x_1}^{\mu}, 
\overline{x_2}^{\mu}, \ldots, \overline{x_p}^{\mu} \}$, 
$(\mu=1, \ldots, p)$, span a
$p$-dimensional vector space. Then, for $r>p$ there are coefficients 
$\alpha_{rl}$ such that 
\begin{equation}
\overline{x_r}^{\mu} = \displaystyle{\sum_{l=1}^p} \alpha_{rl} \overline{x_l}^{\mu}
\end{equation}
for all $\mu=1, \ldots, p$. It follows that every column $(A_i)_{jr}$, 
($i$ fixed,
$j$ a running index of $V_i$ and $r$ a fixed number larger than $p$) is a
linear combination of the first $p$ columns of $(A_i)_{js}$ ($i$ fixed, 
$j$ a running index of $V_i$ and $s$ smaller than or equal to $p$).
Consequently, the matrix $A_i$ has a vanishing determinant, and is not 
invertible.
Therefore, in case the average squared deviation (\ref{variance})
would vanish, the unique solution (\ref{solutionwij}) would not exist.
The fact that for vanishing variances $\sigma_j$ our 
set of equations for the final weights is under-determined has been mentioned 
already in the introduction, in the text under equation
(\ref{Ourequation}).

In \cite{Hee00} the occurrence of the average squared deviation 
(\ref{variance}) has been overlooked. This 
enabled the authors to solve the Master Equation (\ref{masterw4})
in the usual way. By means of the so-called Gauss-Seidel procedure they 
obtained a modified version of the usual pseudo-inverse solution for the 
connections, rather than the expression (\ref{solutionwij}).

\section{Intermediate values for the weights}
\label{intermediatesec}
Since our approach was simply based on the assumption of convergence of 
the $\langle w_{ij} \rangle_n$ for $n\rightarrow \infty$, 
we had no knowledge of the intermediate values  
of the weights $\langle w_{ij} \rangle_n$ for finite $n$. 
However, we can predict the evolution of the weights 
through an iterative procedure.
If we repeat the derivation in section \ref{solutionsec}, starting from 
(\ref{masterw4}) in stead of (\ref{masterw5}), we find
\begin{eqnarray}
\fl \langle w_{ij} \rangle_{n+1}  =  \langle w_{ij} \rangle_n + 
\frac{\eta_i}{p}
\displaystyle{\sum_{\mu=1}^p}\kappa(2\overline{x_i}^{\mu}-1)
\overline{x_j}^{\mu} \nonumber \\
 - \frac{\eta_i}{p} \displaystyle{\sum_{\mu=1}^p}
\Bigl(\displaystyle{\sum_{k=1}^N}
\langle w_{ik} \rangle_n \overline{x_k}^{\mu} - \theta_i\Bigr)
\overline{x_j}^{\mu}
- \eta_i \sigma_j^2 \langle w_{ij} \rangle_n \ \ \ j \in V_i
\label{recurs}
\end{eqnarray}
In the limit $n \rightarrow \infty$, equation (\ref{recurs}) implies 
(\ref{nlarge2}), provided that the weights $\langle w_{ij} \rangle_n$ converge.

\begin{figure}
\begin{center}
\epsfig{file=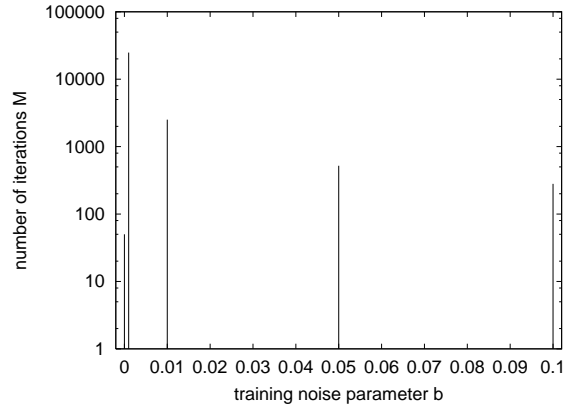, angle=0, width=0.6\textwidth}
\caption{\emph{Convergence time as a function of the training noise parameter.}
The number of iterations $M$ of the recursive formula (\ref{recurs}) is plotted
as a function of the training noise parameter $b$. 
The number $M$ is determined with the help of the criterion that it is the 
smallest value of $n$ for which the condition 
$\sum_j |\langle w_{ij}\rangle_n-\langle w_{ij} \rangle_{\infty}|<0.01$ is 
satisfied for a fixed value of $i$. The network consists of $N=128$ neurons, 
the number of patterns
is $p=16$. Furthermore, we chose $\theta_i =0$ for all $i$ and $\kappa=1$.
The learning rate is $\eta_i=0.25$ for all $i$. 
The smaller $b$, the more iterations are needed to obtain the exact final
value $\langle w_{ij} \rangle_{\infty}$. However, for $b=0$, $M$ drops to 
$50$. In this case, of course,
$\langle w_{ij} \rangle_{\infty} = w_{ij}^{PI}$.}
\label{convergfigure}
\end{center}
\end{figure}

Using the relation (\ref{recurs}), one can find, by numerical iteration, 
the quantities 
$\langle w_{ij} \rangle_n$ for any $n$, given the starting values
$\langle w_{ij} \rangle_0$. Hence, we can verify numerically that the 
$\langle w_{ij} \rangle_n$ are independent of these starting values.
Moreover, one can study the 
convergence of the learning procedure. In order to do so, one must make a 
particular choice for 
the probability distribution $p^{\mu}(\vec{x})$, which, up to now, was left
unspecified. For our choice [see
(\ref{pjmub})], this distribution will
depend on a so-called noise parameter $b$ $(0 \leq b \leq 1)$, such that 
$\overline{x_j}^{\mu} = \xi_j^{\mu}$
and $\sigma_j^2=0$ if the noise parameter $b$ vanishes $(b=0)$. 
Through the parameter $b$ we can tune the
amount of noise during the learning process.
Numerical calculations show that the 
$\langle w_{ij} \rangle_n$ do indeed converge in the limit 
$n\rightarrow \infty$, for arbitrary $b$, including $b=0$, if $\eta_i$ is 
small enough. 
Interestingly, convergence times to the final values (\ref{solutionwij}) 
diverge for a decreasing noise parameter $b$ (\emph{i.e.}, $b \rightarrow 0$),
but the time of convergence drops to a small value if
$b=0$ (see figure \ref{convergfigure}), indicating that 
something peculiar happens in this limit. 

In other words, if one demands existence
of the solution (\ref{solutionwij}), one may choose $b$ arbitrarily small, 
but not zero, and convergence to the solution is faster for larger values of
$b$. 
If one puts $b$ to zero in the iterative application of (\ref{recurs}), one 
observes rapid convergence of 
the weights, to the pseudo-inverse values (\ref{PI}). Maybe 
surprisingly, these values have no continuous relation with the values for 
finite $b$, 
despite the fact that the expressions (\ref{recurs}), the difference 
equations that determine the weights $w_{ij}$, do depend 
continuously on the $\sigma^2_j$, and, hence [see equation (\ref{sigmab}) 
below], on $b$. In view of the difference in the solutions
$\langle w_{ij} \rangle_{\infty}$ for the cases $b=0$ 
[eq. (\ref{solutionwij})] versus $b\neq0$ [eq. (\ref{PI})], 
this is obvious: there cannot be a continuous relationship between them, 
since the pseudo-inverse solution (\ref{PI}) depends on all the initial
values $w_{ij}(0)$, whereas our solution (\ref{solutionwij}) is 
independent of the initial values of the weights $w_{ij}(0)$ for $j \in V_i$. 

In the next section we investigate whether our final solution for the weights
corresponds to the storage of patterns in a stable way.

\section{Stability}\label{stabilitysection}
It is well-known that a neural net with fixed weights $w_{ij}$ 
(in our case this will be after the learning phase) and 
deterministic neuron dynamics evolves, in the course of time, to limit cycles 
of 
finite length $n$ $(n=1,2, \ldots, 2^N)$. Cycles with $n=1$, or fixed points, 
are of
particular interest in neural network theory. If a pattern $\vec{\xi}^{\mu}$ 
is a fixed point of the dynamics of a neural net for a given set of
weights $w_{ij}$, the stability coefficients
(\ref{stabcoef}) or (\ref{stabiliteit})
are positive for all $i$, in which case the system remains in the pattern 
$\vec{\xi}^{\mu}$, \emph{i.e.}, the system is stable \cite{KinOpp91}. 
Besides the fact that $\gamma_i(\vec{\xi}^{\mu})$ is a measure for the size 
of the basin of 
attraction of fixed point $\vec{\xi}^{\mu}$ \cite{Gar88}, it 
is
plausible that it is also a 
measure that determines to what extent the network state $\vec{x}$ remains
in a neighborhood of $\vec{\xi}^{\mu}$ when the deterministic evolution 
of this neuron state $\vec{x}$ is
replaced by a stochastic version of this evolution. 
In order
to get an idea of the effectiveness of the learning process discussed in 
the preceding section, we will therefore consider the expectation value of 
the stability coefficients (\ref{stabcoef}) in the limit $n\rightarrow \infty$:
\begin{equation}
\langle \gamma_i(\vec{\xi}^{\mu}) \rangle_{\infty} = (2\xi_i^{\mu}-1)
(\displaystyle{\sum_{j=1}^N} 
\langle w_{ij} \rangle_{\infty} \xi_j^{\mu} - \theta_i)
\label{gammaksi2}
\end{equation}
Once a set of patterns $\{ \vec{\xi}^{\mu} \}$ is known, these quantities can 
be explicitly calculated with the help of the expression (\ref{solutionwij}).
In this section we will attempt to derive, analytically, an approximation of 
(\ref{gammaksi2}) by 
averaging over sets of patterns $\{ \vec{\xi}^{\mu} \}$ with given mean 
activity $a$. 
If, however, we would calculate the average of 
$\langle w_{ij} \rangle_{\infty}$ over these
patterns directly, we would lose all dependence on neuron indices $j$ 
and pattern 
indices $\mu$, such that the correlations with the $\xi_j^{\mu}$ would 
disappear. We therefore have to take a different route.

An approximation for the expectation value (\ref{gammaksi2}) is
\begin{equation}
\langle \gamma_i(\vec{\xi}^{\mu}) \rangle_{\infty} \approx
(2\xi_i^{\mu}-1)( \langle \overline{h_i}^{\mu} \rangle_{\infty} - \theta_i)
\label{gammaksi3}
\end{equation}
where $\overline{h_i}^{\mu}$ is the membrane potential of neuron $i$ 
averaged with respect to $p^{\mu}(\vec{x})$:
\begin{equation}
\overline{h_i}^{\mu} = \displaystyle{\sum_{j=1}^N} w_{ij} \overline{x_j}^{\mu}
\label{potentiaalmu1}
\end{equation}
In fact, the approximation (\ref{gammaksi3}) would be exact if 
$\overline{x_j}^{\mu}$ would be
equal to $\xi^{\mu}_j$ for all $j$, \emph{i.e.}, in the limit that the 
probability function is such that $\overline{x_j}^{\mu}$ equals $\xi_j^{\mu}$,
for all $j$.
The average potential occurring in (\ref{gammaksi3}) can be found from 
(\ref{nlarge2}). Indeed,
multiplying by $\overline{x_j}^{\mu}$ and summing with respect to 
$j \in V_i$, we find from this equation:
\begin{equation}
\fl \displaystyle{\sum_{j \in V_i}} \overline{x_j}^{\mu} \langle w_{ij} 
\rangle_{\infty}
p\sigma_j^2 = \displaystyle{\sum_{\nu=1}^p}\Bigl[ \kappa
(2\overline{x_i}^{\nu} -1)
- (\displaystyle{\sum_{k=1}^N} \langle w_{ik} \rangle_{\infty} 
\overline{x_k}^{\nu} 
- \theta_i)\Bigr] \displaystyle{\sum_{j \in V_i}}\overline{x_j}^{\nu}
\overline{x_j}^{\mu} 
\label{nlarge5}
\end{equation}
where we also used (\ref{variance}). 

The average square deviation $\sigma_j^2$ occurring in this equation depends
on the neuron $j$. In this article we will consider the case in which all
neurons have the same standard deviation $\sigma_j$
\begin{equation}
\sigma_j  =  \sigma \qquad j=1, \ldots, N \label{jindep1}
\end{equation}
\emph{i.e.},
the probability $p^{\mu}_j(x_j)$ is supposed to be 
such that the uncertainty to find 
neuron $j$ in a state $\xi_j^{\mu}$ is the same for all neurons of the 
neural net. 
Using the identity
\begin{equation}
\displaystyle{\sum_{j \in V_i}} \overline{x_j}^{\mu} \langle w_{ij} 
\rangle_{\infty} p\sigma^2 =
\Bigl(\displaystyle{\sum_{j=1}^N} \overline{x_j}^{\mu} 
\langle w_{ij} \rangle_{\infty} - \displaystyle{\sum_{j \in V_i^C}} 
\overline{x_j}^{\mu} \langle w_{ij} \rangle_{\infty}  \Bigr)p\sigma^2
\label{replace}
\end{equation}
we find from (\ref{nlarge5})
\begin{equation}
\fl \Bigl( \langle \overline{h_i}^{\mu} \rangle_{\infty} - 
\displaystyle{\sum_{j \in V_i^C}}
\langle w_{ij} \rangle_0 \overline{x_j}^{\mu}\Bigr) p\sigma^2 = 
\displaystyle{\sum_{\nu=1}^p} [\kappa(2\overline{x_i}^{\nu} -1) + \theta_i - 
\langle \overline{h_i}^{\nu} \rangle_{\infty}] \displaystyle{\sum_{j \in V_i}}
\overline{x_j}^{\nu} \overline{x_j}^{\mu}
\label{nlarge6}
\end{equation}
where we used $\langle w_{ij} \rangle_0 = \langle w_{ij} \rangle_{\infty}$
for $j \in V_i^C$. An alternative form for (\ref{nlarge6}) reads
\begin{equation}
\displaystyle{\sum_{\nu=1}^p} [p\sigma^2\1 + C_i]^{\mu \nu} \langle \overline{h_i}^{\nu}
\rangle_{\infty} = \displaystyle{\sum_{\nu=1}^p} f_i^{\nu}C_i^{\nu \mu} + g_i^{\mu}
\label{nlarge7}
\end{equation}
where $\1$ is the $p\times p$ unit matrix and where $C_i^{\mu \nu}$, the 
correlation matrix for averaged neuron states, is defined by
\begin{equation}
C_i^{\nu \mu} := \displaystyle{\sum_{j \in V_i}} \overline{x_j}^{\nu} \overline{x_j}^{\mu}
\label{Cmunu}
\end{equation}
Furthermore, we abbreviated
\begin{eqnarray}
f_i^{\mu}  =  \kappa(2\overline{x_i}^{\mu} -1) + \theta_i \label{deff}
\\
g_i^{\mu}  =  p \sigma^2 \displaystyle{\sum_{j \in V_i^C}} \langle w_{ij} \rangle_0 
\overline{x_j}^{\mu}  \label{defg}
\end{eqnarray}
Multiplying both sides of the matrix equation (\ref{nlarge7}) by the inverse 
of the 
(symmetric) matrix occurring on its left-hand side we obtain the solution
\begin{equation}
\langle \overline{h_i}^{\mu} \rangle_{\infty} =
\displaystyle{\sum_{\nu, \lambda=1}^p} f_i^{\nu} C_i^{\nu \lambda} 
[(p\sigma^2 \1 + C_i)^{-1}]^{\lambda \mu} + \displaystyle{\sum_{\nu=1}^p} 
g_i^{\nu}[(p\sigma^2 \1 + C_i)^{-1}]^{\nu \mu}
\label{himu}
\end{equation}

Once a particular probability distribution $p^{\mu}(\vec{x})$ of patterns 
centered around $\vec{\xi}^{\mu}$ is given, we can evaluate $f_i^{\nu}$ and 
$C^{\mu \nu}$, and, hence, via (\ref{himu}), the 
average stability coefficient (\ref{gammaksi3}).

In contrast to our expression (\ref{solutionwij}) for the 
expectation value of the final value of the weights
$\langle w_{ij} \rangle_{\infty}$, the result for the 
expected average potential 
$\langle \overline{h_i}^{\mu} \rangle_{\infty} = \sum_j \langle w_{ij} \rangle_{\infty}\overline{x_j}^{\mu} $
does exist for vanishing $\sigma$. 
This is clear, already, from (\ref{himu}), in which the existence of the
inverse $(p\sigma^2 \1 + C_i)^{-1}$ does not depend on the presence of 
the extra term $p\sigma^2\1$ as long as the average patterns 
$\overline{\vec{x}}^{\mu}$ are linearly independent, since then 
$C_i^{\mu \nu}$ is invertible.
Using (\ref{solutionwij}) and (\ref{deff}), and assuming 
$\langle w_{ij} \rangle_0 =0$ for all $j \in V_i^C$ we may write the average 
potential
$\langle \overline{h_i}^{\mu} \rangle_{\infty}$ as
\begin{equation}
\langle \overline{h_i}^{\mu} \rangle_{\infty} =
\displaystyle{\sum_{\nu}\sum_{k,j \in V_i}} (D_i + A_i)_{jk}^{-1}
\overline{x_j}^{\mu} \overline{x_k}^{\nu} f_i^{\nu}
\label{hisubst}
\end{equation}
Comparing this to (\ref{himu}) with (\ref{gimu}), we obtain the identity
\begin{equation}
\displaystyle{\sum_{k,j \in V_i}} (D_i + A_i)_{jk}^{-1}
\overline{x_j}^{\mu} \overline{x_k}^{\nu} =
\displaystyle{\sum_{\lambda}} C_i^{\mu \lambda}[(p\sigma^2\1 + C_i)^{-1}]^
{\lambda \nu}
\end{equation}
Hence, though the matrix $(D_i + A_i)^{-1}$ occurring in (\ref{hisubst}) 
itself does not exist for $b=0$, the above combination 
clearly does: it reduces to $\delta^{\mu \nu}$, as we see from 
the right hand side for $\sigma=0$, implying that in the limit of vanishing 
noise
\begin{equation}
\langle \overline{h_i}^{\mu} \rangle_{\infty} = f_i^{\mu}
\end{equation}
which is already clear from (\ref{himu}) and is equivalent 
to ---the average of--- eq. (\ref{PIequations}).
Thus, although the values of the weights themselves do not have a continuous 
relation with the values corresponding to the pseudo-inverse solution, 
the average values for the membrane potentials, and, therefore, of the 
stability coefficients, do.

In the following we suppose that $w_{ij}=0$ for all $j \in V_i^C$.  
This corresponds to a neural net in which all existing connections are of 
adaptable strength and the only connections with constant strength are the
non-existing connections. 
For $j \in V_i^C$, we then have $w_{ij}(0)=0$, and, hence,
$\langle w_{ij} \rangle_0 = 0$, implying that
\begin{equation}
g_i^{\mu} = 0
\label{gimu}
\end{equation}
Let us choose the probability distribution
\begin{equation}
p_j^{\mu}(x_j) = (1-b)\delta_{x_j, \xi_j^{\mu}} + b\delta_{x_j,1-\xi_j^{\mu}}
\qquad (j=1,\ldots,N)
\label{pjmub}
\end{equation}
which fulfills (\ref{imunorm}), and from which $p^{\mu}(\vec{x})$ 
follows by the prescription (\ref{pmufactor}).
The noise parameter $b$ is a probability ($0 \leq b \leq 1$).  
More specifically, for given $\mu$ $(\mu=1,2,\ldots,p)$, $1-b$ is 
the probability that the 
activity $x_j$ of neuron $j$ equals that of the pattern $\xi_j^{\mu}$.
We suppose that $b$ is small compared to unity. As follows from (\ref{pjmub}),
the noise parameter $b$ is
related to the width of the distribution of input patterns around each 
pattern $\vec{\xi^{\mu}}$. 
We can immediately 
calculate the average neuron state (\ref{average2}) associated with the 
distribution 
(\ref{pjmub})
\begin{equation}
\overline{x_j}^{\mu}(\xi_j^{\mu}) = (1-b)\delta_{1,\xi_j^{\mu}} + 
b\delta_{1,1-\xi_j^{\mu}}
\label{xjavmu}
\end{equation}
the coefficient (\ref{deff})
\begin{equation}
f_i^{\mu} = (2\xi_i^{\mu}-1)(1-2b)\kappa + \theta_i
\label{fimub}
\end{equation}
as well as the average squared deviation (\ref{variance})
\begin{equation}
\sigma^2 = b(1-b)
\label{sigmab}
\end{equation}
The fact that $\sigma^2$ is $j$-independent is a consequence of the 
particular choice (\ref{pjmub}) for $p_j^{\mu}(x_j)$, \emph{i.e.}, of
the fact that all neurons $j$ are supposed to have the same uncertainty to be 
in state $\xi_j^{\mu}$. Hence, the supposition
(\ref{jindep1}) is satisfied.

Let us suppose that the probability that $\xi_j^{\mu}=1$ is $a$, for each $j$
independent of any other neuron index $k$, and, hence, that the probability 
that $\xi_j^{\mu}=0$ is $(1-a)$, for all of the 
patterns $\vec{\xi}^1, \ldots, \vec{\xi}^p$.
We can now use this to arrive at an 
estimate value for the average potential of neuron $i$, eq. 
(\ref{himu}), which is exact in the limit of 
$p \rightarrow \infty$, for $\alpha=p/N$ fixed, and smaller than $1$.
From (\ref{Cmunu}) we find, for $\mu \neq \nu$,
\begin{eqnarray}
\fl C_i^{\mu \nu} \approx  \displaystyle{\sum_{j \in V_i}} \bigl\{ a^2 
\overline{x_j}^{\mu}(1) \overline{x_j}^{\nu}(1) +
a(1-a)\overline{x_j}^{\mu}(1) \overline{x_j}^{\nu}(0) + \nonumber \\
 (1-a)a\overline{x_j}^{\mu}(0) \overline{x_j}^{\nu}(1) + 
(1-a)^2\overline{x_j}^{\mu}(0) \overline{x_j}^{\nu}(0) \bigr\}
\label{Cmunuu}
\end{eqnarray}
while for $\mu=\nu$ we get
\begin{equation}
C_i^{\mu \nu} \approx 
\displaystyle{\sum_{j \in V_i}}\{a\overline{x_j}^{\mu}(1)^2
+ (1-a)\overline{x_j}^{\mu}(0)^2\}
\label{Cmunue}
\end{equation}
Defining the dilution $d$ as the average fraction of neurons from which an 
arbitrary neuron does not have an incoming connection, each neuron has
on the average $N(1-d)$ incoming connections.
Hence, using (\ref{xjavmu}), we find from (\ref{Cmunuu}) and (\ref{Cmunue})
\begin{eqnarray}
\fl C_i^{\mu \nu} \approx N(1-d)\{a(1-a)(1-2b)^2\delta^{\mu \nu} \nonumber \\
 + [a^2(1-b)^2 + 2ab(1-a)(1-b) + (1-a)^2b^2]\}
\label{Cmunu2}
\end{eqnarray}
We thus have achieved that the correlation matrix
$C_i^{\mu \nu}$ for an $N$ neuron net has been expressed in parameters
typical for the network, namely the dilution $d$, the mean activity $a$ 
and the noise $b$. An alternative way to write (\ref{Cmunu2}) is
\begin{equation}
C_i^{\mu \nu} \approx l\delta^{\mu \nu} + m 
\label{Cmunu3}
\end{equation}
where $l$ and $m$ are shorthand  for combinations of the typical network 
parameters $a$, $b$ and $d$ 
\begin{eqnarray}
l & := & N(1-d)a(1-a)(1-2b)^2 \nonumber \\
m & := & N(1-d)[a^2(1-b)^2 + 2ab(1-a)(1-b) + (1-a)^2b^2]
\label{deflm}
\end{eqnarray}
With (\ref{Cmunu3}), the matrix occurring in (\ref{himu}) can be cast into 
the form
\begin{equation}
(p\sigma^2\1 + C_i)^{\mu \nu} \approx (p\sigma^2 + l)\delta^{\mu \nu} + m
\label{iplusC}
\end{equation}
The inverse of a $p$-dimensional matrix $A$ with elements
$A^{\mu \nu} = x\delta^{\mu \nu} + y$ is given by the matrix $A^{-1}$ with 
elements
\begin{equation}
(A^{-1})^{\mu \nu} = [\delta^{\mu \nu}(x+py) - y]/x(x+py)
\label{Ainvers}
\end{equation}
From (\ref{Cmunu3}), and (\ref{Ainvers}) applied to (\ref{iplusC}), we find
\begin{eqnarray}
\displaystyle{\sum_{\lambda}}C_i^{\nu \lambda}[(p\sigma^2 \1 + C_i)^{-1}]
^{\lambda \mu} \approx
\frac{
l[l+p(\sigma^2 +m)]\delta^{\mu \nu}
+ mp\sigma^2 }
{(l+p\sigma^2)[l+ p(\sigma^2+m)]}
\label{CmunuCinvers}
\end{eqnarray}
Substituting this result, together with (\ref{gimu}), into the expression 
(\ref{himu}), yields for the average potential of neuron $i$ in pattern $\mu$ 
the expression
\begin{eqnarray}
\langle \overline{h_i}^{\mu} \rangle_{\infty} \approx 
\frac{
\{l[l+p(\sigma^2 + m)]+mp\sigma^2 \}f_i^{\mu} + mp\sigma^2 
\Sigma_{\nu \neq \mu}f_i^{\nu}}
{(l+p\sigma^2)[l+p(\sigma^2 + m)]}
\label{himufinal}
\end{eqnarray}
The sum over the $f_i^{\nu}$ occurring in this expression can be calculated 
with (\ref{fimub}), 
\begin{equation}
\displaystyle{\sum_{\nu \neq \mu}} f_i^{\nu} \approx (p-1)[(2a-1)(1-2b)\kappa +
\theta_i]
\label{sumfi}
\end{equation}
where we used that the average value of the $p-1$ neuron activities 
$\xi_i^{\mu}$ can be approximated by $a$, the average activity of the net.
We can now write down the final result for the stability parameters
(\ref{gammaksi3}), which is a function of 
the network parameters $d$, $a$ and $b$, the number of patterns $p$, 
the number of neurons of the net $N$, and the neuron properties $\kappa$
and $\theta_i$:
\begin{eqnarray}
\fl \langle \gamma_i(\xi_i^{\mu}) \rangle_{\infty} 
\approx  
\frac{
\{l[l+p(\sigma^2 + m)]+mp\sigma^2 \}[(1-2b)\kappa + \theta_i(2\xi_i^{\mu}-1)]}
{(l+p\sigma^2)[l+p(\sigma^2 + m)]} \nonumber \\
+ \frac{
mp\sigma^2(p-1)[(2a-1)(1-2b)\kappa +
\theta_i](2\xi_i^{\mu}-1)}{(l+p\sigma^2)[l+p(\sigma^2 + m)]} 
-\theta_i(2\xi_i^{\mu}-1)
\label{finalabdpN}
\end{eqnarray}
Note that for $\sigma^2=0$ we immediately recover 
$\langle \overline{h_i}^{\mu} \rangle_{\infty} = f_i^{\mu}$ and 
$\langle \gamma_i(\xi_i^{\mu}) \rangle_{\infty} = \kappa$, as we should, 
from the equations (\ref{himufinal}) and (\ref{finalabdpN}) respectively.

The final average stability coefficient of 
neuron $i$ takes two different values respectively, depending on whether
$\xi_i^{\mu} = 1$ or $\xi_i^{\mu}=0$.

\begin{figure}
\begin{center}
\epsfig{file=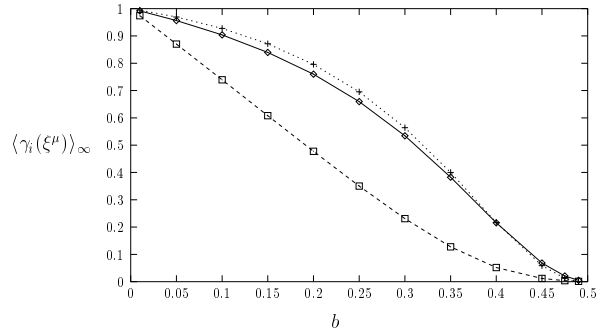, width=0.6\textwidth}
\caption{\emph{Average stability as a function of the training noise parameter}
$b$. The average, over all neurons $i$ $(i=1,2,\ldots, 256)$, and all 
patterns $\vec{\xi}^{\mu}$ $(\mu=1,\ldots,32)$, of a ---diluted--- neural
net (dilution $d=0.2$), of the stability coefficients 
$\gamma_i(\xi_i^{\mu})$, with threshold potentials $\theta_i = 0$ for all $i$,
as a function of the noise parameter $b$. The
curves with symbols ($+$) and ($\opendiamond$), for $\xi_i^{\mu}=+1$ or
$\xi_i^{\mu}=0$, respectively, correspond to a numerical 
simulation, and should be compared to the curve with symbol 
($\opensquare$), obtained from the approximative expression (\ref{finalabdpN}).
The approximation itself is rather poor. However, it gives an indication, 
as the various curves show. 
Note that the appearance of one single curve in the approximated case is due 
to the choice $a=0.5$ and $\theta_i=0$.}
\label{figuur2}
\end{center}
\end{figure}

In figure (\ref{figuur2}) we plotted this quantity for a chosen average 
activity $a=0.5$, as a function of $b$. 
It is clear that the stability 
coefficient can be expected to remain positive. 
In the same figure we plotted
the actual values of $\langle \gamma_i(\xi_i^{\mu}) \rangle_{\infty}$,
as obtained by choosing randomly a set of patterns $\{ \vec{\xi}^{\mu} \}$
with given mean activity $a=0.5$, and using (\ref{gammaksi2}) with
(\ref{solutionwij}) for the calculation.

The difference between the curves is evident, and indicates that we must be 
careful not to overestimate the accuracy of our result as an indication 
for $\langle \gamma_i(\xi_i^{\mu}) \rangle_{\infty}$. In fact, in a large 
region, the 
storage of undisturbed patterns is better than our 
estimate suggests by a factor $2$ to $3$, as can be concluded from the figure.
With this in mind, we may assume that after 
the noisy learning process, the patterns $\vec{\xi}^{\mu}$ are indeed
fixed points under the deterministic network dynamics for a small 
noise parameter $b$.

\section{Retrieval and basins of attraction}\label{basinssection}
\begin{figure}
\begin{center}
\epsfig{file=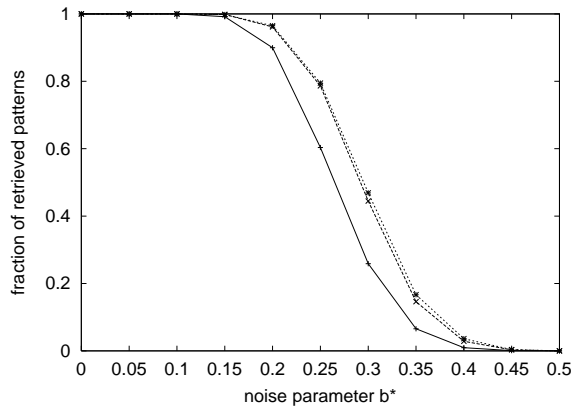, width = 0.6\textwidth}
\caption{\emph{Fraction of retrieved patterns for networks trained with various values of the training noise parameter} $b$, \emph{as a function of the noise} $b^{\star}$
\emph{in the patterns to be retrieved}. 
The average fraction of retrieved patterns for a network of $N=128$ neurons
with thresholds $\theta_i=0$. The values of the dilution and the 
activity are $d=0.2$ and $a=0.5$. 
The network has been trained with $p=32$ patterns.
Starting from an initial pattern with noise $b^{\star}$, the network evolves 
under parallel deterministic dynamics. The pattern $\vec{\xi}^{\mu}$ 
is said to be retrieved if the network state overlap with this pattern,
$m^{\mu} := N^{-1}\sum_{i=1}^N (2\xi_i^{\mu}-1)(2x_i -1)$, is equal to $1$. 
At most $10$ dynamical time steps are applied for every initial pattern.
The various curves $(+)$, $(\times)$ and $(\ast)$
correspond to $b=0$, $b=0.05$ and $b=0.1$ 
respectively. The value of $b$ for which the size of the basins of attraction
is maximal ---for the values of $b$ chosen in this figure--- is $b=0.1$. 
The curve with $b=0$ (noiseless training) lies below the curves 
corresponding to learning with noise.}
\label{basins}
\end{center}
\end{figure}

In this section we address the question what happens to the average size of 
the basins of attraction if noiseless learning (training parameter $b=0$) is 
compared to noisy learning ($b\neq 0$). After the network has been trained with
patterns $\vec{x}$ with noise $b \ (b\neq0)$, we numerically check the 
retrieval 
capacity of the net by presenting patterns with noise $b^{\star}$, \emph{i.e.},
patterns chosen according to a probability distribution of the form 
(\ref{pjmub}), in which $b$ has been replaced by $b^{\star}$. 
The presented patterns evolve under deterministic parallel dynamics
$x_i(n+1) = \Theta(h_i(n)-\theta_i)$.  
The attempt to retrieve a pattern is successful if the network state 
$\vec{x}$ runs into a fixed point equal to 
the clean, undistorted pattern $\vec{\xi}^{\mu}$ of which a noisy 
version was the initial state.
The result can
be read off from figure \ref{basins}. Since the curves obtained via noisy 
learning lie above the curve with noiseless learning, the basins of 
attraction are, apparently, enlarged in the presence of noise during the 
learning stage. 

The result is in agreement with earlier studies by
Gardner et al \cite{GSW89}, and Wong \& Sherrington 
\cite{WS90a}, \cite{WS90b}, \cite{WS93}. 

In \cite{GSW89}, like in our case, noise is added to patterns
during a training stage. However, the algorithm is of a different kind, 
because it includes
an error-mask, \emph{i.e.}, the weights are updated if and only if,
upon presentation of a noisy pattern during the learning stage, the membrane
potential $h_i$ has the wrong sign. In this way, if the learning algorithm
converges, retrieval of patterns for which the amount of noise is equal
to that of the training patterns, is guaranteed. 

In \cite{WS90a}, \cite{WS90b} and \cite{WS93}, various retrieval properties 
of a neural network are discussed. It is argued that
optimizing (by finding the optimal weights) the overlap of a noisy pattern 
with its corresponding training representative after one retrieval step is, 
in fact, a way of noisy training \cite{WS90a}. 
The optimal network is sought for via a
replica-calculation that minimizes a cost-function, thus optimizing the first 
step retrieval. No explicit learning rule is used in these articles.
An explicit expression for the final values of the weights is not given.
 
Our approach is different from those discussed above, in the sense that
we start from an explicit learning rule, which is biologically acceptable: 
it is 
derived from the principle that energy cost for synaptic adaptation is 
minimal \cite{Hee99}; it is a function of local variables;
it does not contain error masks; neurons are assumed to be noisy. Though 
in our network the basins of attraction are not optimal
(it was not our goal to optimize the basins of attraction), we do have an 
explicit
learning algorithm as well as a final expression for the expectation value of
weights.

\section{Conclusion}\label{conclusionsec}

We have shown that learning with noise leads to final values for the
weights $w_{ij}$ which are different from those found in the
corresponding situation without noise. Surprisingly, the solution for the 
values of the weights $w_{ij}$ of a noisy system in the limit of vanishing 
noise, does \emph{not} converge to the values of the solution of the system
without noise.

Moreover, in a system without noise the values of the final
weights depend on the initial values of all weights,
whereas in a noisy system the initial values of the
changing weights, the $w_{ij}$ for $j \in V_i$, are wiped out in the course of
time. 

Our noisy trained networks have larger basins of attraction than noiselessly
trained networks. This is in agreement
with earlier findings in the literature. The exact dependence of
the retrieval properties on various parameters, 
such as the mean activity $a$ and the memory load $\alpha = p/N$ is still to
be elucidated.

\section*{Acknowledgment}
The authors are indebted to Wouter Kager for carefully reading this manuscript
and suggesting some improvements.
\\
\\

\end{document}